\documentclass[aps,prl,twocolumn,showpacs,superscriptaddress,groupedaddress]{revtex4-1}  
\usepackage{graphicx}  
\usepackage{dcolumn}   
\usepackage{bm}        
\usepackage{amssymb}   

\begin{document}

%
%

\title{Co-detection of micro seismic activity as early warning of gravitational slope failure}
%
%
%

\author{J\'erome Faillettaz}
\affiliation{3G, UZH, University of Z\"urich, Z\"urich, Switzerland}
\author{Martin Funk}
\affiliation{VAW, ETHZ, Z\"urich, Switzerland}
\author{Jan Beutel}
\affiliation{TIKZ, ETHZ, Z\"urich, Switzerland}
\author{Andreas Vieli}
\affiliation{3G, UZH, University of Z\"urich, Z\"urich, Switzerland}

%
%
%

\begin{abstract}
We developed a new strategy for Disaster Risk Reduction for gravitational slope failure: 
We propose a simple method for real-time early warning of gravity-driven failures that considers and exploits both the heterogeneity of natural media and characteristics of acoustic emissions attenuation. This method capitalizes on co-detection of elastic waves emanating from micro-cracks by a network of multiple and spatially distributed sensors. Event co-detection is considered as surrogate for large event size with more frequent co-detected events marking imminence of catastrophic failure. In this study we apply this method to a steep rock glacier / debris slope and  demonstrate the potential of this simple strategy for real world cases, i.e. at slope scale. 
This low cost, robust and autonomous system provides a well adapted alternative/complementary solution for Early Warning Systems.
\end{abstract}
\maketitle
\section{Introduction}

Slope and rock instabilities due to permafrost degradation, rockfalls, landslides, snow avalanches or avalanching glacier instabilities are common in high mountain areas. These gravity-driven rupture phenomena occurring in natural heterogeneous media are rare, but have a potential to cause major disasters, especially when they are at the origin of a chain of processes involving other materials such as snow (snow avalanche), water (flood) and/or debris (mudflow) \citep{Gill&Malamud2014}. They potentially endanger mountain communities or real estate development  and are at the origin of huge human fatalities and economic costs \citep{Petley&al2005,Sidle2006}. In the context of climate warming, degradation of permafrost is expected to further promote slope destabilization in high mountains and thus increase the occurrence of such natural disasters \citep{Gruber&al2004}. Because of the potential magnitude of such catastrophic phenomena, a reliable forecasting combined with a timely evacuation of the endangered areas is often the most effective way to cope with such natural hazards. However, the nonlinear nature of geological material failure hampered by inherent heterogeneity, unknown initial mechanical state, and complex load application (rainfall, temperature, etc.) hinder predictability. 

Slope stability assessment is often based on monitoring of the temporal evolution of external parameters such as geometry, surface displacement (or surface velocity) as well as on the observation of external forcing such as meteorological conditions (e.g., rainfall duration and intensity, temperature, wind, snow accumulation,...). On the basis of a theoretical/modeling study, \cite{Faillettaz&al2016} recently proposed a new method to investigate natural slope stability based on continuous monitoring and interpretation of seismic waves generated by the potential instability - i.e. an internal parameter. 
This method capitalizes on both heterogeneity and attenuation properties of natural media for developing a new strategy for early warning systems: As heterogeneous materials breaks gradually, with their weakest parts breaking first, they produce precursory “micro-cracks” with associated elastic waves traveling in the material. Therefore the monitoring of such micro-seismic activity offers valuable information concerning the progression of damage and imminence of global failure \citep{Michlmayr&al2012}. Such monitoring are providing new insights into the imminence of break-off and in some cases it has been applied to natural gravity-driven instabilities such as cliff collapse \citep{Amitrano&al2005}, slope instabilities \citep{Dixon&al2003,Kolesnikov&al2003,Dixon&Spriggs2007} or failure in snow pack \citep{vanHerwijnen&Schweizer2011,Reiweger&al2015}.  
 However, as elastic waves travel in the material, their amplitudes decay with distance from the source. Due to attenuation of propagating acoustic/seismic signals (elastic waves), an event (i.e., a crack formation in the material) may also be observed and recorded differently by an acoustic/seismic sensor depending on its location. Theoretical considerations based on simple numerical modeling suggest that, although statistical properties of attenuated signals amplitude could lead to misleading results, detecting emergence of large events announcing impeding failure (precursors) is possible even with attenuated signals \citep{Faillettaz&al2016}. 
 It requires a network of (seismic/acoustic) sensors on a potential unstable slope and and the detection of events in real time. Real-time processing of measured events that are detected concurrently on more than one sensor (co-detected) enables then to easily access their initial magnitude as well as their approximate initial location. This simple method based on co-detection of elastic waves traveling through natural media provides a simple means to access characteristics and temporal evolution of surrogate variables linked to hillslope damage and mechanical state. For this method to function, temporal synchronization between sensors must be sufficiently accurate to reliably classify events detected simultaneously by multiple sensors, therefore the sensor network needs to be precisely synchronized. Preliminary application  to acoustic emissions during failure of snow samples at lab scale has confirmed the potential usefulness of co-detection as indicator for imminent failure. 

To demonstrate the application potential of this simple strategy for Early Warning Systems to real cases, i.e. at slope scale, we designed and built an experimental system composed of a network of six seismic sensors wired to a data acquisition unit, ensuring an interleaved sampling time synchronization between sensors. This experimental setup was installed and tested on the steep tongue of the Dirru rock glacier, a location where small scale slope instabilities were highly probable. Note that the steep slope is composed of an highly heterogeneous material consisting of a mixture of ice, rock, fine sediment, air and water. 
In this study, we show the first results of the analysis of the seismic activity generated by the steep tongue during summer 2017. Thanks to a meteorological station located closed to the rock glacier and L1 Differential GPS unit on the rock glacier \citep{wirz&al2013}, we were able to investigate the relation between seismic activity, surface displacement and external forcing (rainfall, temperature). Using additional webcam images with a time interval of 30 minutes, we identified three small scale failure events (of approximately 10 cubic meters each) and analyze the associated number and temporal evolution of co-detection prior to to failure. This co-detection analysis showed typical patterns of precursory events prior failure, demonstrating thus the potential of this method for real world applications in early warning. Moreover, this seismic method provides new insights on the rock glacier dynamics, especially the short term peaks of velocity in relation to external forcing.

The motivations of this study are twofold:
First, it aims at testing the applicability of the co-detection method at the slope scale and thus at demonstrating its  application potential in the context of natural slope stability assessment.
 Second, as our experiment was deployed on a fast moving rock glacier, we had the opportunity to investigate, for the first time, the seismic activity emitted by the glacier tongue and its link to complex rock glacier dynamics.
 
The paper is organized as follow: After describing the study site and the experimental setup, we performed the analysis of the co-detection method and demonstrate its potential applicability to early warning of gravity driven geofailure. Comparing results with all available data, ranging from surface displacement to meteorological data, complex rock glaciers dynamics is discussed in the light of these new observations.

\begin{figure*}
\includegraphics[width=\textwidth]{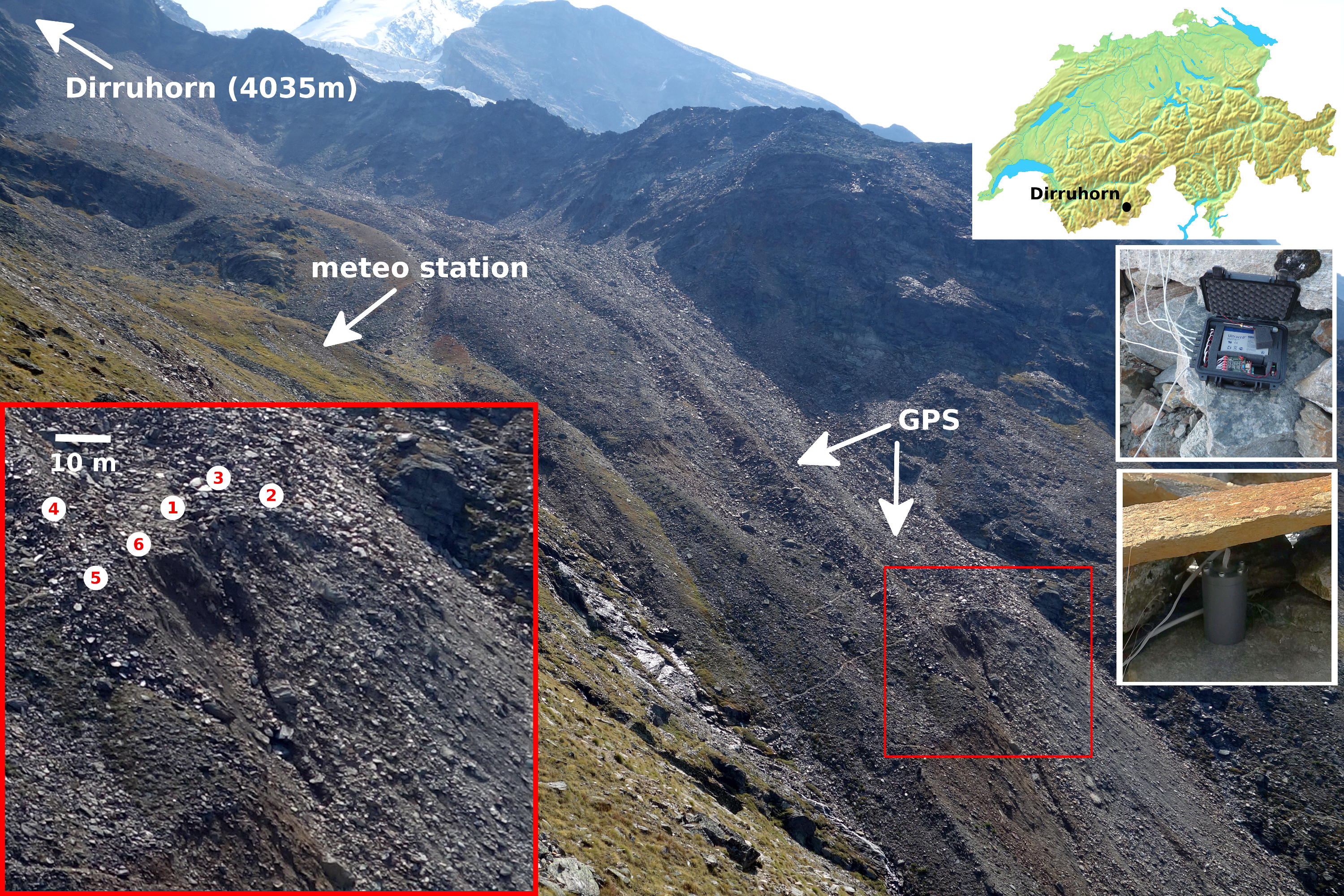}
\caption{\label{general}General view of the Dirru rock glacier. White arrows indicate the location of the GPS and meteo stations installed on the rock glacier and analyzed in this study. Bottom left inset: Location and associated number of each sensor installed near the steep tongue of the Dirru rock glacier. Top right inset: General location of Diru glacier. Middle right inset: View of the central data acquisition unit, each sensor being wired to this unit. Bottom right inset: View of a sensor installed on the field with a large rock sheltering it.  }
\end{figure*}

\section{Study site and experimental setup}

\subsection{Study site}
The study site is located in the area of the Dirruhorn in the Matter Valley, above Herbriggen/Randa, Switzerland. The mainly westerly exposed slopes range from 2600 to 4000 m a.s.l.. Permafrost is abundant in this area \citep{Delaloye&al2010}. The field area includes various cryosphere-related slope movements: e. g. exceptionally fast and potentially dangerous rock glaciers moving up to 10 m/a \citep{Delaloye&al2010}.
The rock glacier Dirru is composed of various lobes and  fronts,  originating  from  different  rock  glacier
generations. The currently active lobe, which is located on the orographic right side of Dirrugrat, has a total length of more than 1 km, is about 60 to 120 m wide, and is approximately 20 m thick \citep{wirz&al2016}. It has a convex profile and slope angles increase from about 15$^o$ in the  upper  part  to  more  than  30$^o$ in  the  lower  part  towards its front. Since the 1970/80s this steep frontal part (tongue) has progressively accelerated and reached surface velocities above 5 m.$\rm a^{-1}$, potentially indicating a phase of destabilization  \citep{Delaloye&al2013}. Its front already collapsed in some parts in the recent past. At this front, water emerges occasionally in spring and summer. Based on past photographs,  it was found that the actual acceleration phase of its frontal part started progressively during 1970s and 1980s and that the origin of the destabilization of the entire rock glacier seems to be older \citep{Delaloye&al2013,wirz&al2016}.

\subsection{Field experiment setup}
	
The seismic experimental setup is composed of six geophones (Ion SM-6, one channel with a natural frequency of 10 Hz) directly wired to a central data acquisition unit (Fig. \ref{general}, right inset), ensuring a good time synchronization.
 Each sensor is also embedded in a waterproof casing specially designed for these sensors (Fig. \ref{general}) . A pre-amplifier was also  installed to mitigate attenuation effects in the 20 meters cables.
A data acquisition unit was built and designed specially for this experiment. It is composed of a mini computer Arduino able to record and store on a SD card signal amplitude of the 6 sensors at a sampling rate of 250 Hz. 

The procedure for recording data is the following: As soon as a signal with an amplitude higher than a preset threshold is detected, the Analog to Digital Converter (ADC) is powered on and data is recorded from all sensors for one second. If, during this period, one of these sensors records an amplitude higher than the preset threshold, the whole array continues to record for another whole second.
If the activity is high, this procedure could result in a single long record. During the monitoring period (11 July - 5 September 2017) the maximum duration of a signal was around 500 seconds.

Besides the highly probable occurrence of failure events during summer, this site was also selected for a pilot experimental study because of the proximity to other concurrent measurements setup during the Xsense I and II projects. During this period, air temperature and precipitations were monitored (from the meteo station installed few hundreds meters from the tongue, see Fig. \ref{general}) along with a webcam that took images from the tongue at a 30 minutes interval (Fig. \ref{general}).
These images provide valuable information on the timing, the location and the rough magnitude of failure events occurring at the the tongue.
Events ranging from single rockfalls/rockslides to large slides were detected. Analyzing the seismic activity during these short periods provides  a unique way to investigate the seismic signature of each event, and thus to characterize the potential precursory seismic signals associated with each event.
During bad weather conditions the webcam images were obscured by fog, but this only occurred a few days during the observation period in summer (<7 days).
Two differential L1-GPS sensors permanently installed on the fast moving part of the rock glacier were also monitoring surface displacement (Fig. \ref{general}).


\section{Results and analysis}

\subsection{General overview of meteorological conditions and rock glacier dynamics}
\begin{figure}
\includegraphics[width=.5\textwidth]{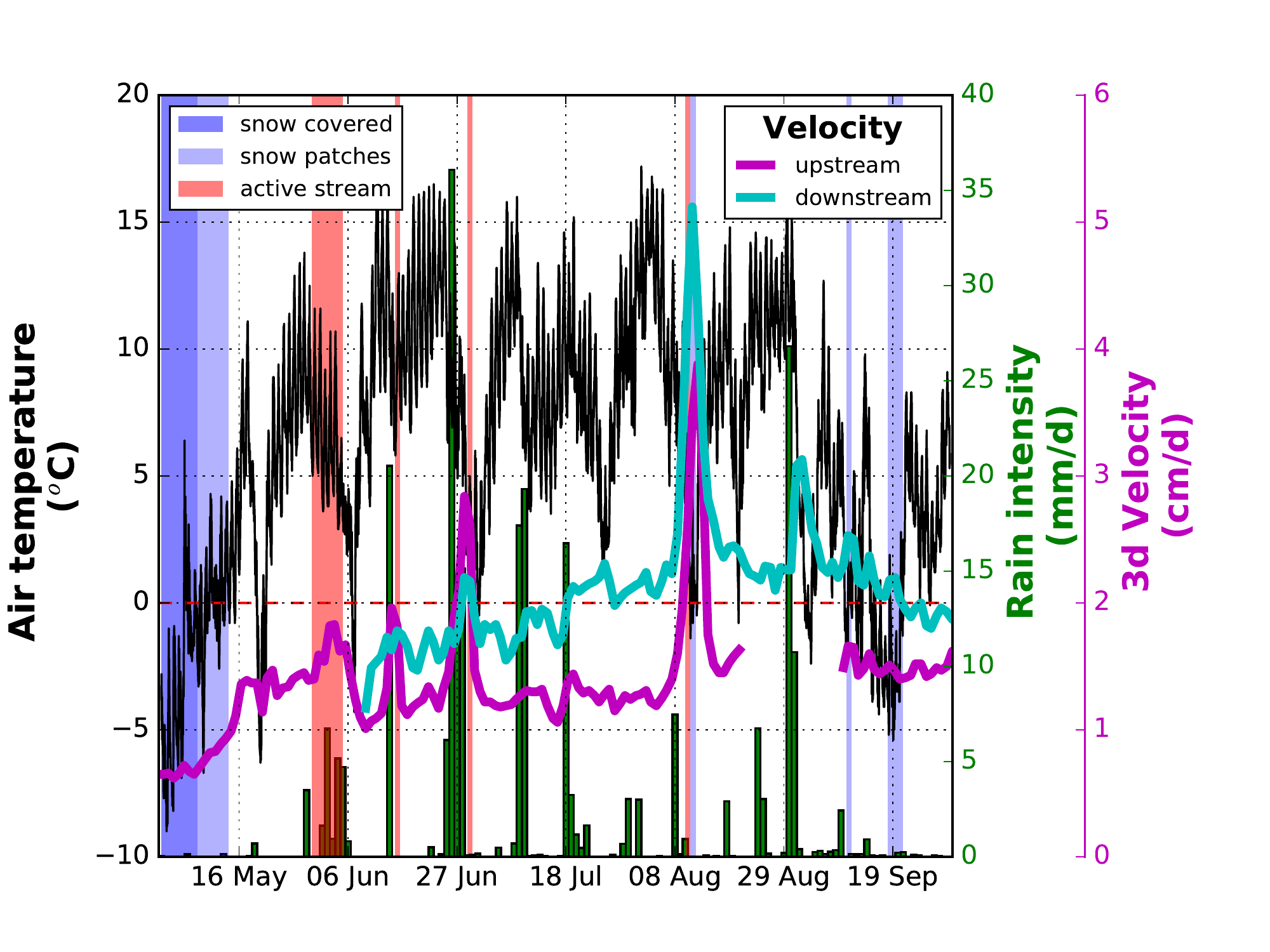}~
\caption{\label{PT} Temperature (black line), precipitations (green bars) and surface velocity (blue and magenta line) of the rock glacier during summer 2017. Upstream and downstream velocities refers to the velocities of two differential L1 GPS stations located in the upper and lower part of the rock glacier tongue, respectively. Red bars in background indicate periods when a stream was spilling out the rock glacier tongue, dark blue period when snow fully covered the rock glacier and light blue periods when snow only partially covered the rock glacier. }
\end{figure}

Fig. \ref{PT} shows temperature, precipitation and surface velocity of the rock glacier at two different locations (Fig. \ref{general}), over the monitoring period in summer 2017. As already observed  for earlier years by \cite{Wirz2016}, rock glacier movement shows a seasonal pattern with an increase starting with the snow melt and reaching maximum flow in late summer/early autumn. In addition to these seasonal variations, short-term peaks in surface velocity are also recorded, in agreement with previous observations \citep{wirz&al2016}. During such peaks, velocity approximately double over a period of a few days to speeds of a few centimeters per days (2 to 5 cm.$\rm d^{-1}$) and drop rapidly to their initial value.
These peaks seem to be related to the presence of large amounts of liquid water within the glacier. Indeed they appear after intense precipitation events or during the snowmelt period. Moreover, a stream spilling out of the tongue, indicating substantial flux of liquid water within the rock glacier, was observed four times in the summer period (May to September) and once during the monitoring period, i.e. 11 July - 5 September 2017 (indicated with a red band on Fig. \ref{PT}). Note that the occurrence of such water outflow is also concomitant to such short-term speed-up events.
 
\begin{figure}
\includegraphics[width=.5\textwidth]{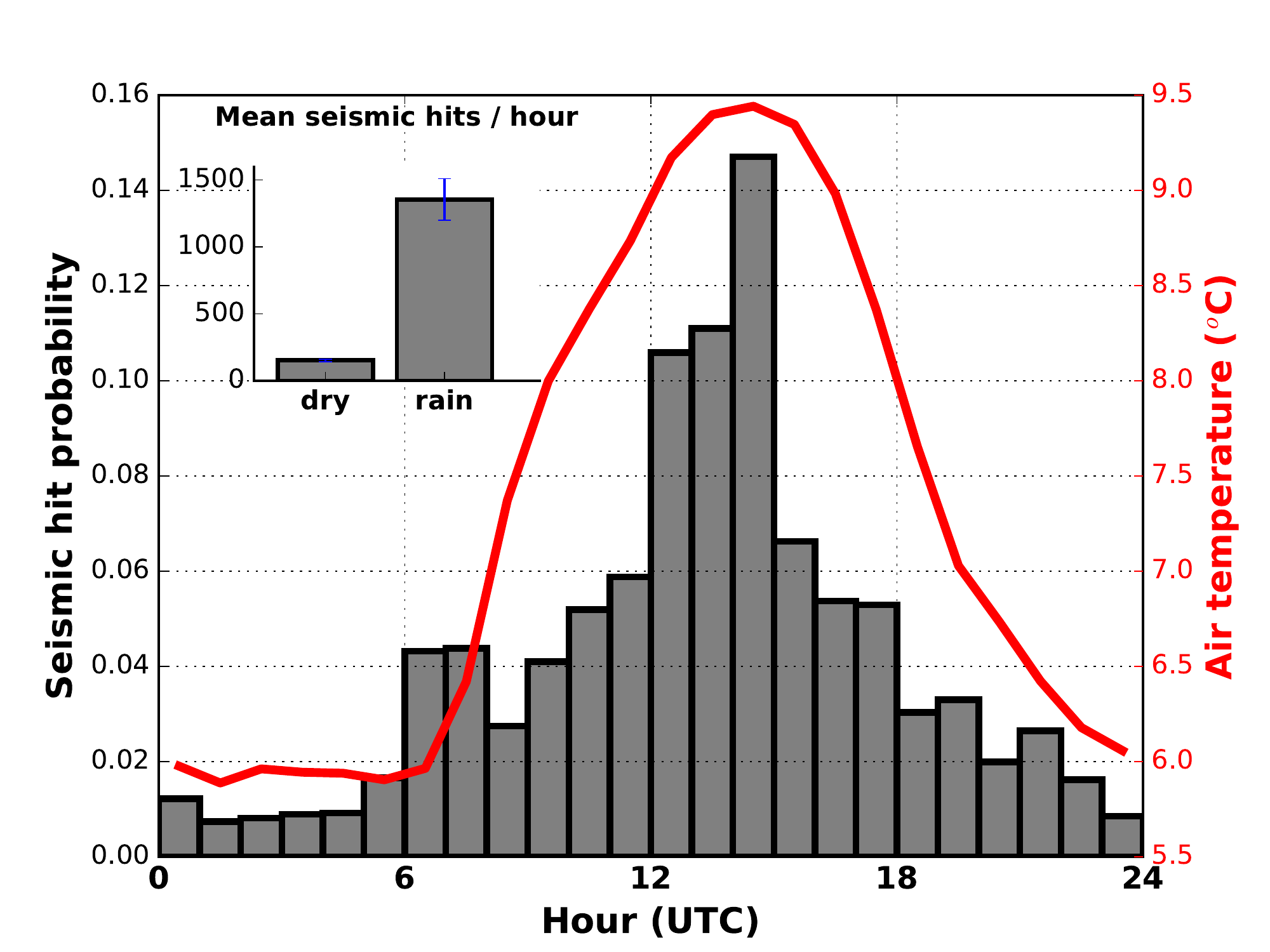}
\caption{\label{seismic} Mean hourly seismic activity (expressed in hits per hour) during the seismic monitoring period (11 July - 5 September 2017). The inset shows the difference in seismic activity between wet (i.e. when liquid precipitations occurred) and dry days.}
\end{figure}

Fig. \ref{seismic} shows the hourly seismic activity (seismic hit probability) emanating from the rock glacier tongue during the monitoring period. Seismic activity shows a clear correlation with air temperature: The number of seismic events is increasing during the day, reaching its maximum concurrently to the maximum in air temperature.
Further, the seismic activity is shown to be clearly higher during periods when liquid precipitations occurred  (Fig. \ref{seismic} inset).
As a result, the seismic activity generated by the steep rock glacier tongue appears to be strongly correlated with both air temperature and the presence of liquid water (rainfall or snowmelt).

\begin{figure*}
\includegraphics[width=\textwidth]{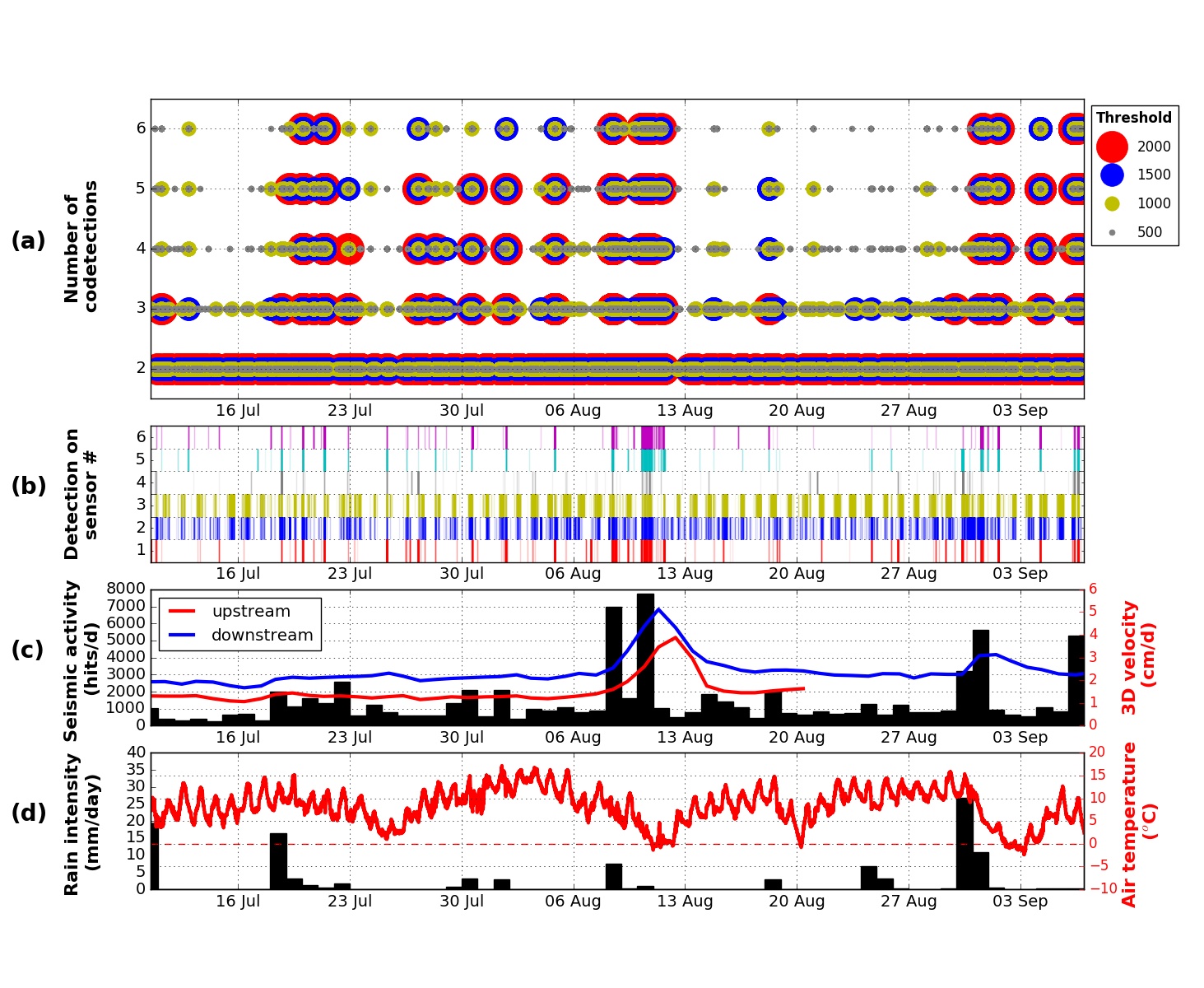}
\caption{\label{cod} (a) Number of co-detections as a function of time using different detection thresholds (colored different sized circles), the larger the threshold, the less sensitive the detection (the numbers given - 500 to 2000 - are arbitrary, without unit), (b) Event detection per sensors number, each vertical line represents a detected seismic event (c) Daily seismic hits (bars) and mean daily velocity from two different GPS locations (blue and red lines), (d) Air temperature (red line) and liquid precipitations (black bars) recorded at the meteorological station located few hundreds of meters from the tongue (see Fig. \ref{general}). The seismic monitoring period ranges from 10 July to 5 September.
}

\end{figure*}
\subsection{Co-detection}
The number of co-detection is defined as the number of sensors that detect an event emanating from the same source (i.e., the signal amplitude is larger than a predefined threshold). 
In practice, we counted the number of sensors detecting a signal within a short period (here 0.1 second), this time window being evaluated according to the sensor spacing and the signal propagation in the medium.
Although the real detection threshold is given by the properties of the sensors and the setup, it could be increased during the post analysis, as the full waveforms of the seismic signals are concurrently recorded and the trigger threshold in the original setup set sufficiently low. 
Fig. \ref{cod} shows (a) the number of co-detection as a function of time using different detection thresholds, the larger the threshold, the less sensitive the detection (the numbers given - 500 to 2000 - are arbitrary, without unit), (b) which sensor is detecting an event, (c) the daily seismic hits, the mean daily velocity from two different locations and (d) the air temperature and precipitation during the monitoring period.

In general, the number of co-detections exhibits similar trend as the seismic activity (total number of seismic events detected by the network, independently of their amplitudes or energy, third panel of Fig. \ref{cod}): During the monitoring period, three periods with high seismic activity and high number of co-detection can be highlighted (17-21 July, 8-11 August and 1-4 September). The initiation of these active seismic phases occurred after wet periods (rainfall event or snow melt event, i.e periods of high air temperatures).
Whereas surface velocity exhibits a slightly increasing trend (except few velocity peaks shortly after or close to enhaced seismic activity), the seismic activity or the number of co-detection show a different temporal variation pattern during the monitoring period and even a calm period (e.g. 13-20 August, Fig. \ref{cod}c), indicating that glacier dynamics and seismic activity are not directly correlated.

As already shown in Fig. \ref{seismic}, a rainfall event (i.e., a direct addition of liquid water on the rock glacier) increases seismic activity at the tongue, but Fig. \ref{cod} shows that the response is not linear: Low precipitation rates are sometimes related to high activity (e.g. 7$\rm^{th}$ August), whereas during large rainfall events only a small increase in seismic activity is recorded (e.g. 17$\rm^{th}$ July).

The sensors 2 and 3, located closed to the steep left-side front, are detecting more seismic events than the others, whereas 
sensor 4, located few tens of meter upstream the front, detects substantially less events. Even if these sensors are not located that far apart (less than 50 meters), the recorded seismic activity is substantially different, demonstrating thus that the attenuation phenomenon has a huge influence on seismic monitoring.
%

\begin{figure*}
\includegraphics[width=.8\textwidth]{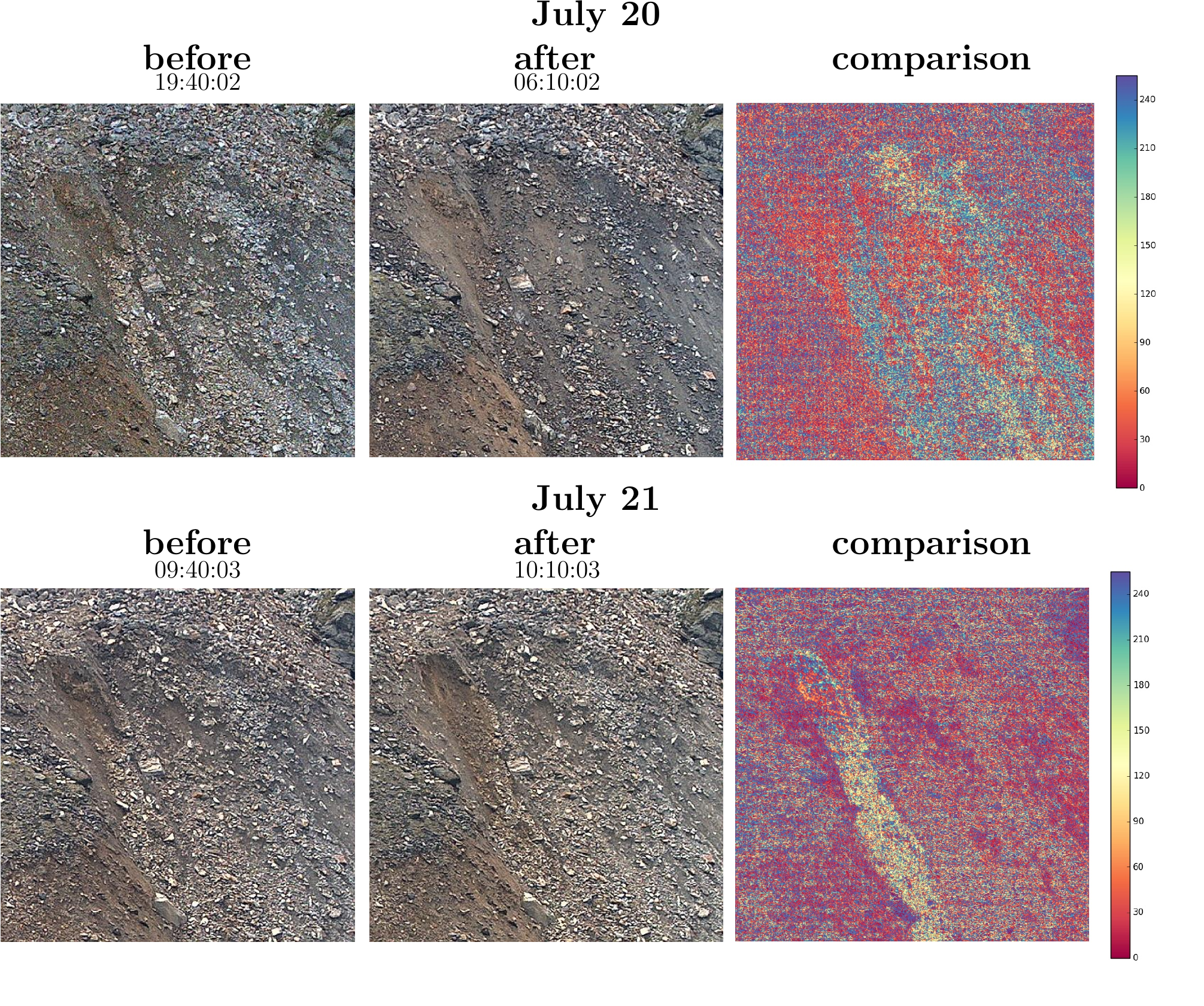}
\caption{\label{comp} Close-up images (800 * 800 pixels) taken from the webcam during the endogenous failures of 20$\rm ^{th}$ July and 21$\rm ^{st}$ July 2017.
First column shows the last exploitable image (with its exact timing) before the associated event, the second column the first exploitable image after the event. The
 third column shows the differences, where yellow and blue colors highlight locations experiencing the larger mismatch between images.}
\end{figure*}
\begin{figure}

\includegraphics[width=.5\textwidth]{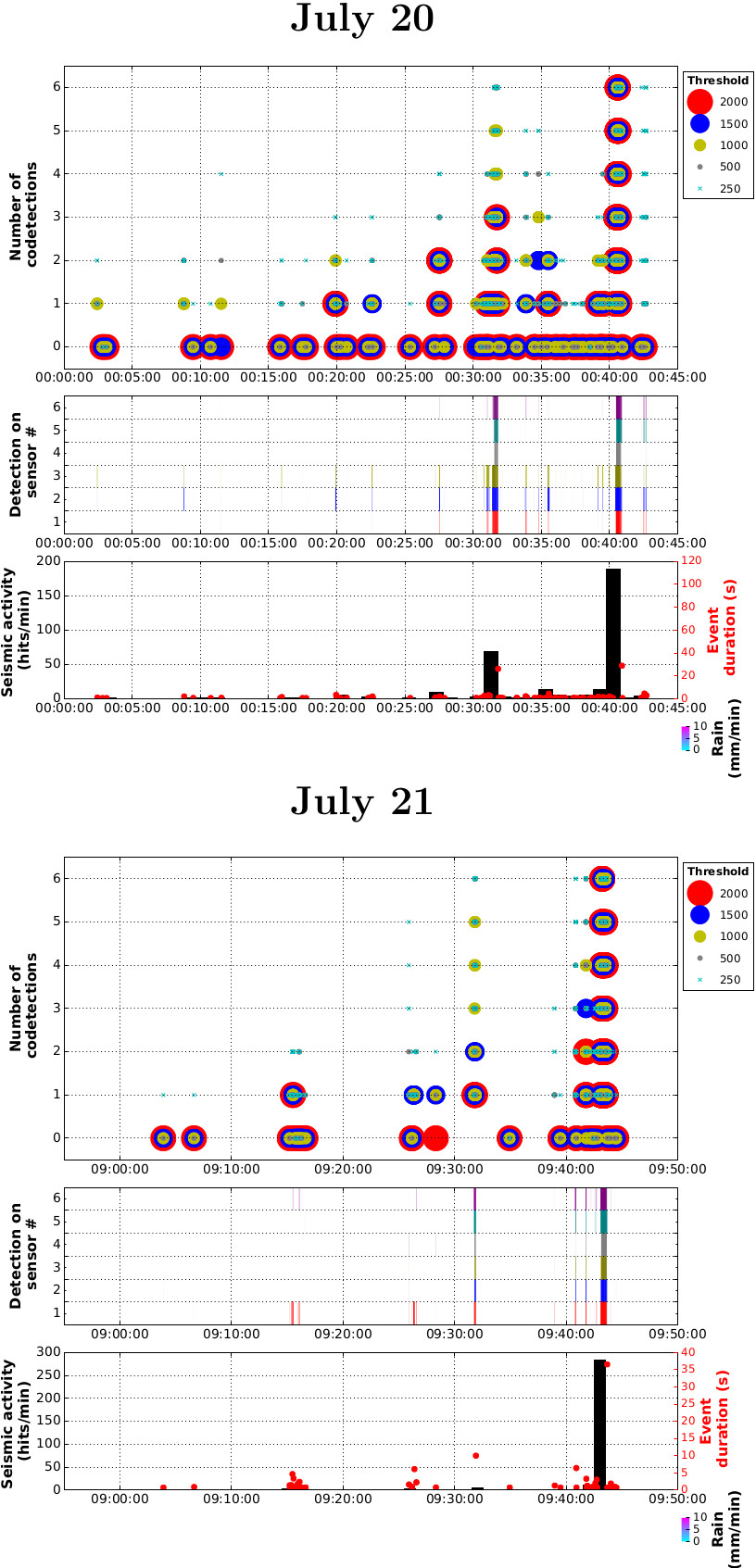}
\caption{\label{endo}  Endogenic failure (top panel: 20 July, bottom: 21 July): Number of co-detection using different thresholds (same arbitrary unit as in Fig. \ref{cod}), their associated detecting sensors, the corresponding seismic activity (and event duration) and the precipitation record during this period.} 

\end{figure}

\begin{figure}[t]
 \noindent\includegraphics[width=.5\textwidth]{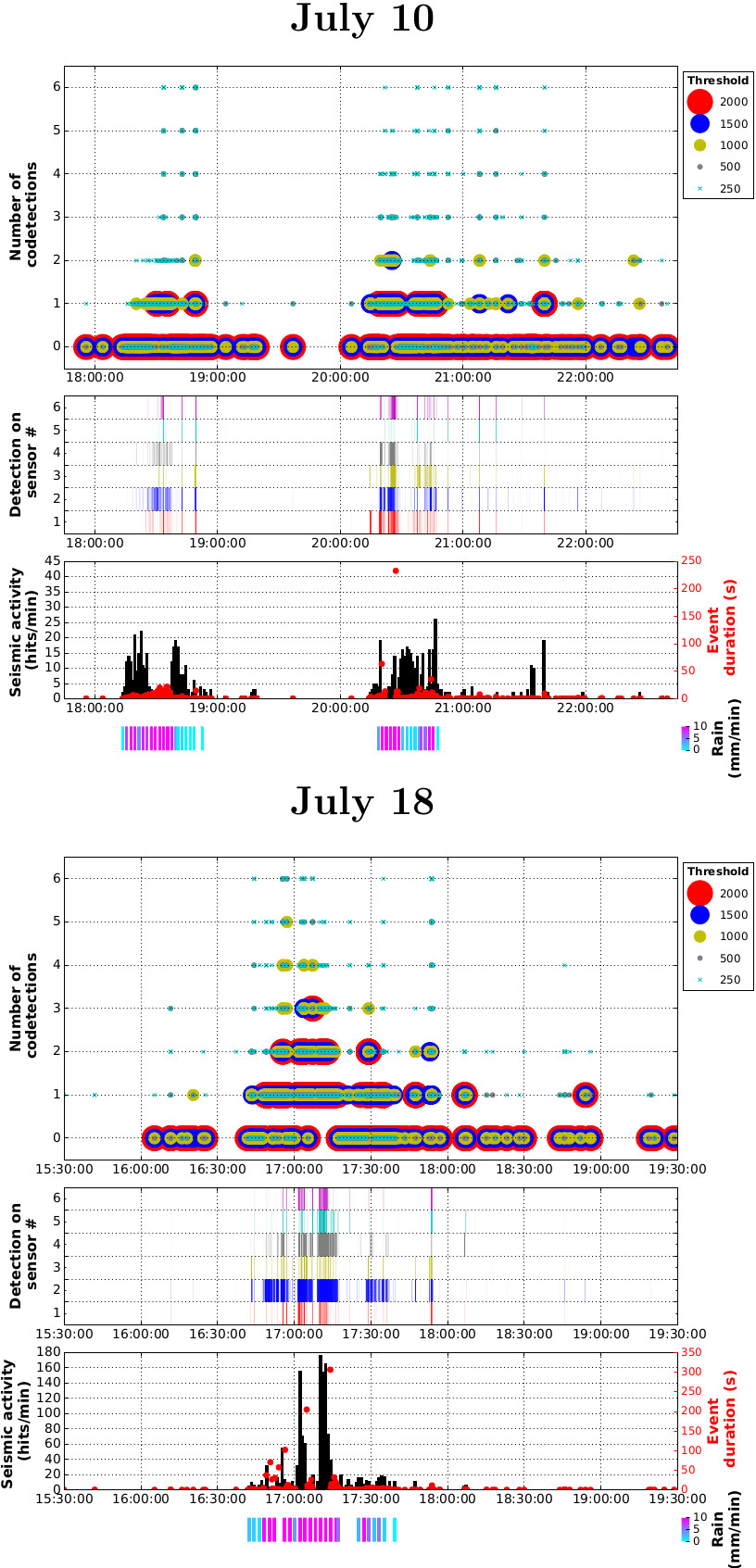}
\caption{\label{exo}  Exogenic failure (top panel: 10 July, bottom: 18 July): Number of co-detection using different thresholds, their associated detecting sensors (same arbitrary unit as in Fig. \ref{cod}), the corresponding seismic activity (and event duration) and the precipitation record during this period. }

\end{figure}
\clearpage
\subsection{Destabilization process and associated seismic precursors:}
In general, slope destabilization can result either from dynamical or quasi-static processes, respectively from a change in the external forcing (i.e. exogenous failure) or from internal changes (i.e., endogenous failure).
Resulting seismic signatures of such types of failure are expected to thus be different. The different type of data from our field experiment allows to identify and analyze both exogenous and endogenous failures during the monitoring period. Failure events were detected using the webcam images with a temporal resolution of 30 minutes (when usable). Results are shown in Fig. \ref{endo} and \ref{exo}.

 

During the monitoring period, we identified two clear failure events corresponding to endogenous (internal) failures (Fig. \ref{endo}): Differences between consecutive usable webcam images show undoubtedly small "landslide"-type events (3-10 $\rm m^3$) occurring at the tongue during dry periods. According to the webcam images, such confirmed debris-slides occurred (i) between July 19 at 19:40 and July 20 at 06:10 (a long time interval because of night)  and (ii) on July 21 between 09:40 and 10:10. During these periods, the recorded seismic activity was low with almost no co-detections except during short periods, i.e. on 20 July at 00:45 and on 21 July at 09:45 (Fig. \ref{comp}). As landslide-type events release high seismic energy generating large seismic waves (e.g. shocks between rolling blocks), these high numbers of co-detection might correspond to the exact timing of the occurrence of the instabilities.
 As no rainfall occurred during and in the 2 days preceding these events, the destabilization was not directly triggered by changes in external forcing, and could thus be attributed to an endogenous failure.
The detailed analysis of the co-detection monitoring of two of these periods is shown on Fig. \ref{endo}. 
These endogenous events exhibit strong similarities: (1) a clear increase in seismic activity and increasing number of co-detection about 45 minutes prior to the failure event (a pattern as expected by \citep{Faillettaz&al2016}), (2) the occurrence of a precursory event 10 to 15 minutes prior the main failure, (3) a strong increase in the number of detection of the sensors located close to the final event, allowing to some degree to locate the final event (event 20 July between sensor 2 and 3, 21 July between 1 and 6 and 31 August near 2).

\begin{figure*}[!]

\includegraphics[width=\textwidth]{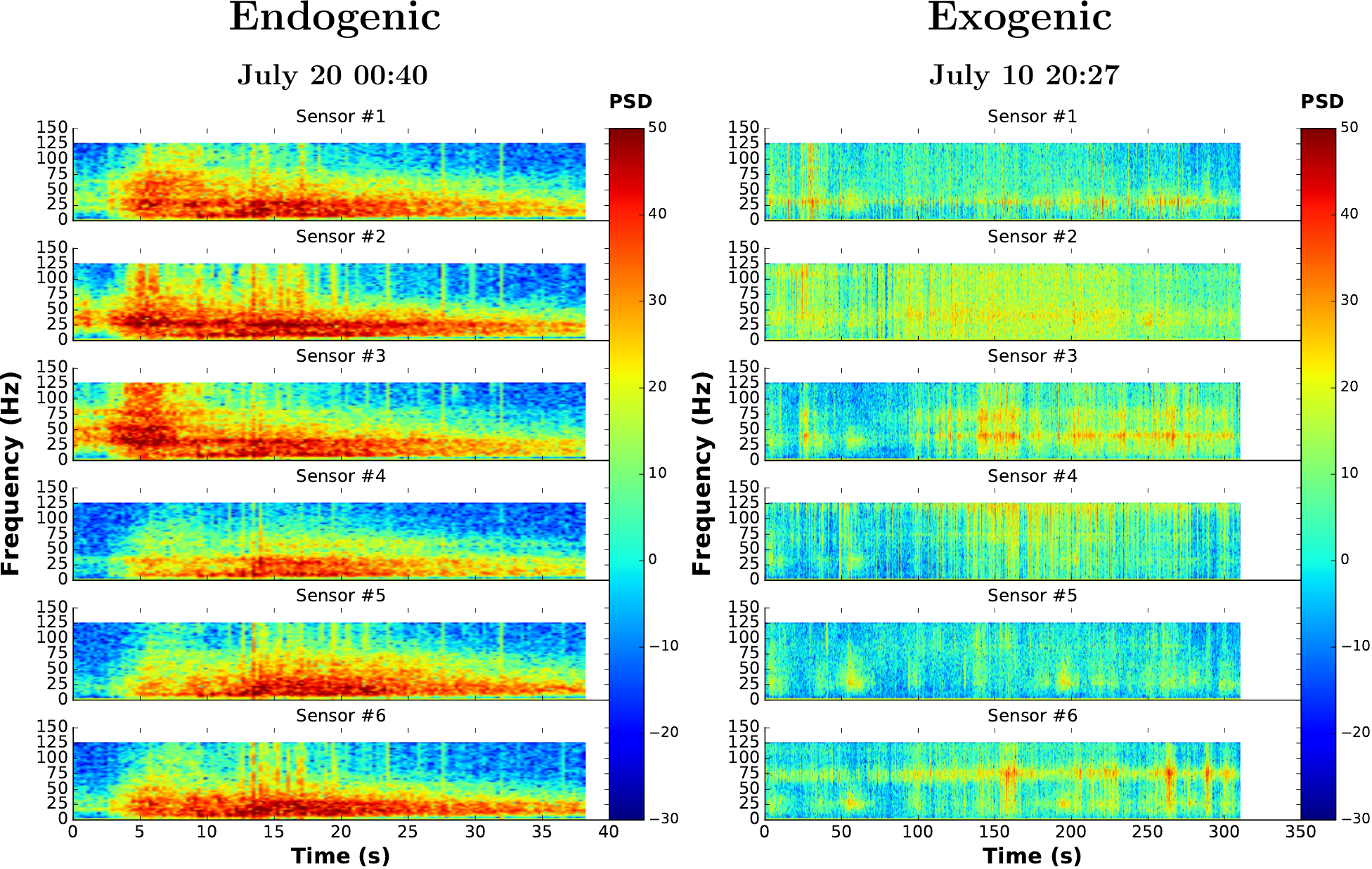}

\caption{\label{spec}  Typical power spectral densities (i.e. spectrograms) associated with endogenic and exogenic events, plotted with the same scale. Frequency domains are colored from red for high power to blue for low power.}

\end{figure*}

Exogenous failures (externally driven) were also identified from the webcam images during large rainfalls. The detailed analysis for two typical events is shown in Fig. \ref{exo}. In such cases, a different seismic activity has been recorded: Although the seismic activity is very high, there is only a few co-detections, indicating  that such seismic events have low amplitudes. In contrast to endogenous failure, no clear precursors can be found. 
Moreover, the analysis of the spectrograms shows a clear difference in the frequency content of these events: 
whereas endogenous failure exhibits a  dominant frequency around 20-40 Hz (highlighted in red in Fig. \ref{spec}), exogenous event are less energetic, with only a few isolated frequency bands containing with substantial energy, apparently linked to each sensor location (Fig. \ref{spec}).

\section{Discussion}
Seismic waves captured by our geophone-network system can be produced by the initiation or propagation of internal cracks, by the landslide event itself but also by surface activity, i.e. small rock sliding and rolling on the steep tongue, or rearrangement of the larger blocks located at the surface of the rock glacier. The direct impact of rainfall on the geophone can also create seismic signals, but, as we sheltered the sensors by large stones (Fig. \ref{general}), we will excluded this process as a potential source of seismic activity.

During periods with external forcing (i.e. rainfall, snow melt periods), it appears that exogenous seismic activity is rather distributed along all the sensors, indicating an homogeneous distribution of seismic events over the rock glacier (Fig. \ref{cod}b). Movements in unconsolidated materials (or over a pre-exisiting failure plan) or progressive melting under large superficial blocks are not expected to produce seismic waves. Moreover,
such exogenic events appear to be less energetic with longer duration (Fig. \ref{spec}) than the endogenous events (Fig. \ref{endo} and \ref{exo}). 
This suggests that the exogenous seismic activity is mainly produced by the sudden rearrangement of the superficial blocks of the rock glacier: As the rock glacier experiences superficial acceleration, the blocks located at the surface can be moved to unstable positions. Rainfall events can then trigger sudden readjustment of blocks, as water lubricates the contacts between the larger blocks and thus reduces friction. 
Of course, blocks located near to the steepest part of the tongue might also slide and roll, thus explaining the slight increase in seismic activity detected close to the tongue.
No precursory signs of failure is thus expected for exogenous failures.

As our co-detection strategy makes possible to separate endogenous from exogenous activity, periods, timing and locations of debris release can be quantified and, hence, rough estimates of debris delivery from the tongue can potentially be derived.
 Such information is needed for debris flow modeling, as the initial volume of unstable debris is a key parameter to model debris flow runout. 

We analyzed different "landslide"-type events based on our new strategy and concurrently analyzed the variations of glacier velocity during this period. In this particular experiment, two endogenous events (on 20 and 21 July) occurred during relative "slow" periods (1-3 cm/d, see Fig. \ref{cod}) and a low seismic activity emanating from the tongue. 
In contrast, the co-detection analysis combining different post analysis thresholds showed a clear increase before each event, thus indicating  that the proposed strategy has for this particular example a better potential application to prediction of failure than seismic activity or even surface displacement.
The co-detection method provides also another metric helping experts to assess slope stability, this metric being related to the ongoing destabilization of the rock glacier.

\begin{table*}
\caption{The different types of failure and their associated behavior}
\centering
\begin{tabular}{l|cccc}
Failure type & Seismic activity & Co-detection number & Precursor & Power spectral density\\
exogenous & high & low & no & low\\
endogenous& high& high + increasing& yes, 10-15 minutes & high\\
\end{tabular}
\end{table*}

In this pilot study  we were able to find precursory signs announcing the impeding failure for small landslides. Moreover, analyzing the spatial distribution of the sensors detecting this precursory seismic activity provides a rough estimate of the location of the potentially unstable zone. 
The existence of precursors to catastrophic failure highly depends on the nature of the rupture process \citep{Faillettaz&Or2015}. For ductile-like rupture, a lot of precursors are expected to occur, suggesting thus a high potential for Early Warning perspectives. In contrast, for brittle-like rupture, even if precursors exist, they are seldom \citep{Faillettaz&Or2015}. In this case, the ongoing destabilization is expected to be more difficult to detect in advance. 
However, the proposed method has clear potential to assess the general type of rupture by studying the effect of external forcing (rainfall for example) on the generated seismic activity. 
As \cite{Faillettaz&Or2015} proposed with their universal global failure criterion (damage weighted stress), a sudden change in external forcing may directly imply an enhanced production of seismic waves for ductile-like failure. In contrast, for brittle-like failure, the external forcing is not expected to produce any additional seismic activity. Studying the seismic response to a change in the external forcing would then offer a direct characterization of the nature of the rupture at stake on a particular slope. In this way, even if no seismic activity is recorded during a change in external forcing, the system might also provide new insights on the nature of the studied instability. "Listening to silence"  in combination with  observing external forcing might also  be as relevant as capturing seismic events.

To really assess efficiently slope stability, a long term monitoring is needed. As every slope is different (composed of different materials, having different external forcing,... ), their behavior will differ. Therefore, the instantaneous seismic activity emanating from the slope does not provide conclusive information for stability assessment purposes. Continuous monitoring of the seismic activity over a long period will allow to establish a reference state enabling then to detect potential changes and trends in behavior, and therefore to estimate/assess the state of stability. 

We demonstrated that the sudden increase of co-detected events is a good indicator of slope destabilization, providing more insights than seismic activity. However, defining a suitable criterion based on the number of co-detections for assessing slope stability and provide early warning perspectives still needs to be determined.  Analyzing concurrently the same set of data (waveforms) using different post-analyzed detection threshold allows to better characterize the size and location of the precursory events. However, it is not clear if such analysis is able to determine such a robust threshold criteria for co-detection,
 the maximum co-detection number, i.e. the number of sensors, being too small (six) to characterize an increase. 
 Moreover, different metrics can be used to define a  criterion suitable for early warning: Such criterion could be based on (i) an absolute number of co-detected events which would be easy to be implement in real-time, but, as every slope is different, such type of criterion might depend on the overall background noise and number and spatial arrangement of the sensors; (ii) on the differences in the temporal evolution of co-detections for different detection thresholds; or on 
(iii) the statistics of "record breaking" events, in the same way as in the mean field model of fracture \cite{Danku&Kun2014}. Records are bursts (i.e. seismic events) which have the largest size since the beginning of the time series, hence their behavior involves extreme values statistics. \cite{Danku&Kun2014} showed that, thanks to such analysis, two regimes of the failure process can be identified, one dominated by the disorder of the material (corresponding to a relative slowdown of the records dynamics) and another dominated by the enhanced triggering of events towards failure (characterized by a temporal acceleration of the record dynamics). Performing such type of co-detection analysis would provide a direct way to assess the time of the failure, even if the initial state is not known.

\section{Conclusions}
In order to demonstrate the application potential of this simple co-detection strategy for Early Warning Systems to real cases, i.e. at slope scale, we designed and built an experimental system composed of a network of six geophones wired to a central recording unit, ensuring thus a perfect time synchronization between the sensors. This experimental setup was installed and tested on the steep tongue of the Dirru rock glacier, a location where small scale slope instabilities were highly likely. To our knowledge, this constitutes the first detailed seismic study on a rock glacier. Note that the steep slope is composed of an highly heterogeneous material resulting from a mixture of ice, rock, fine sediment, air and water. In this study, we present the first results and analysis of the seismic activity generated by the steep tongue during summer 2017. Using additional data from a meteorological station and GPS located on the rock glacier, we were able to investigate the relation between seismic activity, surface displacement and external forcing (rainfall, temperature). Using an additional webcam taking images at a time interval of 30 minutes, we could identify three small-scale failure events (of approximately 10 cubic meters) and analyzed the associated number of co-detected events prior to failure. This detailed analysis allowed us to detect typical patterns of precursory events prior slide events, demonstrating the potential of this method for a real word applications. Moreover, such a seismic method provides new insights on the rock glacier dynamics, especially on the short term peaks of velocity in relation with external forcing.
Additionally, as this simple strategy filters out the small seismic events (generally produced by exogenous failure), only the information relevant for slope stability assessment is delivered and analyzed.

As a next step we propose to develop low-cost tightly integrated sensors that can communicate the relevant seismic data in a wireless manner and in real time with a sufficient time synchronization (less than 0.1 s). 
As the principle of this method is quite general and is virtually applicable to all gravity-driven instabilities, potential applications are numerous ranging from natural hazard prevention and warning of snow avalanches, rockfall, landslides, debris flow, moraine stability, glacier break-of to glacier lake outburst,.... 
Thanks to its simplicity and its robustness, this new strategy would (a) reduce the amount of data to be processed (as only the precise detection time is needed, not the waveform), (b) simplify data analysis and thus enable on-site real time analysis, (c) provide low energy monitoring solution and (d) have low production cost. This new system - that tracks the in situ evolution of a potential unstable slope in real time- would provide a low cost, robust and simple alternative to the existing  Early Warning Systems. 

%

\textit{Acknowledgments} -  
This work was partly supported by the project X-Sense2 funded by the nano-tera.ch.
The authors thank Diego Wasser for handling the technical development of this pilot field experiment and Pr. Reynald Delaloye (Univerty of Fribourg, Switzerland) for providing the webcam images.

\bibliographystyle{agufull08}


\end{document}